\documentclass[
preprint,showpacs,preprintnumbers,amsmath]{revtex4}

\usepackage{amsmath,amssymb,amscd,amsbsy,amsgen,amsopn,amstext,
amsxtra}

\usepackage{color}

\usepackage[dvips]{graphicx}

\begin{document}

\title{{Uncertainty principle, Shannon-Nyquist sampling\\ and beyond}}
\author{{Kazuo Fujikawa$^{\dagger}$}, {Mo-Lin Ge$^*$}, {Yu-Long Liu$^*$}
and {Qing Zhao$^*$}}
\affiliation {$^\dagger$Mathematical Physics Laboratory, RIKEN Nishina Center,
Wako 351-0198, Japan}
\affiliation{$^*$School of Physics, Beijing Institute of Technology, Haidian District, Beijing 100081, P.R. China}


\begin{abstract}
Donoho and Stark have shown that a precise deterministic recovery of missing information contained in a time interval shorter than the time-frequency uncertainty limit is possible. We analyze this signal recovery mechanism from a physics point of view and show that the well-known Shannon-Nyquist sampling theorem, which is fundamental in signal processing, also uses essentially the same  mechanism. The uncertainty relation in the context of information theory, which is based on Fourier analysis, provides a criterion to distinguish Shannon-Nyquist sampling from compressed sensing. A new signal recovery formula, which is analogous to Donoho-Stark formula, is given using the idea of Shannon-Nyquist sampling; in this formulation, the smearing of information below the uncertainty limit as well as the recovery of  information with specified bandwidth take place. We also discuss the recovery of states from the domain below the uncertainty limit of coordinate and momentum in quantum mechanics and show that in principle the state recovery  works by assuming ideal measurement procedures. The recovery of the lost information in the sub-uncertainty domain  means that the loss of information in such a small domain is not fatal, which is in accord with our common understanding of the uncertainty principle, although its precise recovery is something we are not used to in quantum mechanics.  
The uncertainty principle provides a universal sampling criterion covering both the classical Shannon-Nyquist sampling theorem and the quantum mechanical measurement.
\end{abstract}


\maketitle

\section{Introduction}
 Donoho and Stark~\cite{donoho} have shown that a precise deterministic recovery of missing information contained in a time interval with a size $T$ shorter than allowed by the time-frequency uncertainty principle $T\geq 1/W$~\cite{heisenberg, kennard, robertson} is possible. Here $W$ stands for the bandwidth and  this specific form of uncertainty relation, which is used in information theory, generally arises in the context of  the conditional measurement in quantum mechanics~\cite{davies}. This idea of deterministic signal recovery was originally discussed at the early stage of
 the developments of compressed sensing. However, as is explained below, this idea is not included in the compressed sensing as presently understood, namely, the recovery of a signal from highly incomplete measurements by utilizing side information such as sparsity~\cite{candes1, donoho2, candes2}.

The idea of the uncertainty principle in quantum mechanics is well-known, but the recovery of information from the domain below the uncertainty limit is something new to physicists. The purpose of the present paper is to analyze this signal recovery mechanism from a physics point of view and show that the well-known Shannon-Nyquist sampling theorem~\cite{nyquist, shannon}, which is fundamental in signal processing, also uses essentially the same  mechanism, namely, the recovery of information from the domain below the uncertainty limit; this connection of Shannon-Nyquist sampling with the uncertainty principle has not been recognized before.
  
  To our knowledge, the present paper is the first to clarify the connection of Donoho-Stark mechanism with Shannon-Nyquist sampling, and the scheme of Donoho and Stark is excluded from the compressed sensing since Shannon-Nyquist sampling is used as a criterion of conventional sensing. A new signal recovery formula which is similar to Donoho-Stark formula using the idea of Shannon-Nyquist sampling is illustrated in eq.(28) below; in this formulation, the smearing of information below the uncertainty limit as well as the recovery of information with specified bandwidth take place.   We also discuss the recovery of information from the domain below the uncertainty limit of coordinate and momentum in quantum mechanics and show that in principle the state recovery  works by assuming ideal measurement procedures.  The uncertainty principle provides a universal sampling criterion characterizing the classical Shannon-Nyquist sampling theorem and the quantum mechanical measurement.
\\

We start with a brief summary of the basic machinery used in the analysis of  Donoho and Stark~\cite{donoho}. We
use a Dirac notation which is directly extended to the case of quantum mechanics. (The use of Dirac notation in classical information theory should not cause confusion. Besides, the Dirac $\delta$-function is commonly used in information theory such as in the discussion of Shannon-Nyquist sampling.)  For the same reason we consider only the $L_{2}$ norm in this paper, although the $L_{1}$ norm is more important in compressed sensing.
We define the projection operators
\begin{eqnarray}
P_{W}=\int_{w_{0}-\frac{1}{2}W}^{w_{0}+\frac{1}{2}W}dw|w\rangle\langle w|,\ \ \
P_{T}=\int_{t_{0}-\frac{1}{2}T}^{t_{0}+\frac{1}{2}T}dt |t\rangle\langle t|,
\end{eqnarray}
using the relations
\begin{eqnarray}
\langle t|t^{\prime}\rangle=\delta(t-t^{\prime}),\ \ \
\langle w|w^{\prime}\rangle=\delta(w-w^{\prime}),\ \ \
\langle t|w\rangle=\exp[-2\pi iwt].
\end{eqnarray}
The projection operator $P_{W}$ is characterized by the frequency domain $[W]\equiv [w_{0}-\frac{1}{2}W, w_{0}+\frac{1}{2}W]$ and its size or bandwidth $|W|=W>0$. Similarly, the projection operator $P_{T}$ is characterized by the time domain $[T]\equiv [t_{0}-\frac{1}{2}T, t_{0}+\frac{1}{2}T]$ and its size or time interval  $|T|=T>0$.
We often use the notation $P_{W}=\int_{[W]}dw|w\rangle\langle w|$ and $P_{T}=\int_{[T]}dt|t\rangle\langle t|$.  We have for a signal represented by $|\psi\rangle$
\begin{eqnarray}
P_{W}\psi(t)&\equiv&\langle t|P_{W}|\psi\rangle=\int_{[W]} dw e^{-2\pi iwt}\hat{\psi}(w),\nonumber\\
P_{T}\psi(t)&\equiv&\langle t|P_{T}|\psi\rangle=\int_{[T]} dt^{\prime}\delta(t-t^{\prime}) \psi(t^{\prime}),
\end{eqnarray}
by noting $\hat{\psi}(w)=\langle w|\psi\rangle$ and $\psi(t)=\langle t|\psi\rangle$. The notations $\psi(t)$ and $\hat{\psi}(w)$ are more common ones in classical information theory. For simplicity, we consider only the intervals $[W]$ and $[T]$. The relation (3)
shows that
\begin{eqnarray}
P_{W}\psi(t)=\int_{[W]} dw e^{-2\pi iwt}\hat{\psi}(w)
=e^{-2\pi iw_{0}t}\int_{[W]_{0}} dw e^{-2\pi iwt}\hat{\psi}(w+w_{0}),\nonumber
\end{eqnarray}
with $[W]_{0}\equiv [-\frac{1}{2}W, +\frac{1}{2}W]$; this formula with a {\em known} factor $e^{-2\pi iw_{0}t}$ is important when we later discuss the relation of Donoho-Stark mechanism to Shannon-Nyquist sampling which is generally defined in terms of $[W]_{0}$. When we compare  the uncertainty relation with Shannon-Nyquist sampling, we simply set $w_{0}=0$ in the following.

We can confirm $P^{2}_{W}=P_{W}$ and $P^{2}_{T}=P_{T}$, and
we have
\begin{eqnarray}
\frac{\langle\psi|P_{W}P_{T}P_{W}|\psi\rangle}{\langle\psi|P_{W}|\psi\rangle}
&=&\frac{\langle\psi|P_{W}(P_{W}P_{T}P_{W})P_{W}|\psi\rangle}{\langle\psi|P_{W}P_{W}|\psi\rangle}
\nonumber\\
&\leq& ||P_{W}P_{T}P_{W}||\leq TW,
\end{eqnarray}
by noting the relation
\begin{eqnarray}
||P_{W}P_{T}||^{2}=||P_{W}P_{T}P_{W}||\leq {\rm Tr}(P_{W}P_{T}P_{W})=WT
\end{eqnarray}
since $P_{W}P_{T}P_{W}$ is positive semidefinite hermitian and thus $||P_{W}P_{T}P_{W}||$ agrees with its largest eigenvalue while ${\rm Tr}(P_{W}P_{T}P_{W})$ counts all its positive eigenvalues. We also used ${\rm Tr}(P_{W}P_{T}P_{W})=\int_{[W]}dw\int_{[T]}dt\langle w|t\rangle\langle t|w\rangle=WT$.
The relations (4) and (5), and the relations where $P_{W}$ and $P_{T}$ are interchanged, correspond to the upper bound to the conditional measurement in the case of quantum mechanics~\cite{davies}; the conditional measurement is defined to measure
$P_{W}$ first and then measure $P_{T}$ for the resulting state in the case of (4).
The use of the modified state for the second measurement, which is related to reduction, is specific to the quantum conditional probability and it is different (particularly in the case of non-commuting operators) from the classical conditional probability given by the Bayes rule.  The relation (4) as an upper bound to the conditional measurement can provide a constraint on the actions of  $P_{W}$ and $P_{T}$ only for $TW< 1$ since the left-hand side of the relation is bounded by unity; for example, $0\leq \langle\psi|P_{W}P_{T}P_{W}|\psi\rangle/\langle\psi|P_{W}P_{W}|\psi\rangle\leq 1$, namely, by the norm of the projection operator $P_{T}$ as is confirmed  using the definitions in (3).

To substantiate the above manipulation in (4), we here give a  direct proof of (4):
\begin{eqnarray}
\frac{\langle\psi|P_{W}P_{T}P_{W}|\psi\rangle}{\langle\psi|P_{W}|\psi\rangle}&=&
\frac{\int_{[T]}dt\int_{[W]}dw \left(e^{-2\pi iwt}\hat{\psi}(w)\right)\int_{[W]}dw^{\prime} \left(e^{-2\pi iw^{\prime}t}\hat{\psi}(w^{\prime})\right)^{\star}}{\int_{[W]}dw |\hat{\psi}(w)|^{2}}\nonumber\\
&\leq& \frac{\int_{[T]}dt\int_{[W]}dw\int_{[W]}dw^{\prime}\frac{1}{2}[|\hat{\psi}(w)|^{2}+|\hat{\psi}(w^{\prime})|^{2}]}{\int_{[W]}dw |\hat{\psi}(w)|^{2}}\nonumber\\
&=& TW.
\end{eqnarray}
This proof shows that the relation (4) is valid for any values of $TW$ as long as
$\int_{[W]}dw |\hat{\psi}(w)|^{2}\neq 0$, and similarly $\int_{[T]}dt |\psi(t)|^{2}\neq 0$. There is  no lower bound to $TW$ except for $TW>0$ unless one adds some extra conditions.

Usually we do not impose the norm such as $L_{2}$ on the time dependence in quantum mechanics. In the context of signal recovery, we understand that the probability smaller than unity in (6) for the case $WT<1$ specifies a ratio of the size of the signal covered by the projection operator $P_{T}$ relative to the entire normalized signal $P_{W}|\psi\rangle/||P_{W}|\psi\rangle||$ measured by the  $L_{2}$ norm.
From the present point of view, the uncertainty principle of Donoho and Stark~\cite{donoho},
\begin{eqnarray}
WT\geq 1
\end{eqnarray}
is based on an {\em additional} assumption of unit probability (or "$\epsilon$-concentrated" condition~\cite{donoho}) which is expressed by, for example,
\begin{eqnarray}
\langle\psi|P_{W}P_{T}P_{W}|\psi\rangle/\langle\psi|P_{W}|\psi\rangle=1.
\end{eqnarray}
This relation (complete measurement) means that  $||P_{T}P_{W}|\psi\rangle||=||P_{W}|\psi\rangle||$ together with $P_{W}P_{W}|\psi\rangle=P_{W}|\psi\rangle$. The uncertainty principle by Donoho and Stark is thus a necessary condition for the satisfactory description of a given signal $|\psi\rangle$ by $P_{W}$ and $P_{T}$, or a condition so that the measurements of  $P_{W}$ and $P_{T}$ are consistently performed for the signal  $|\psi\rangle$. The quantum mechanical uncertainty relation for $p$ and $x$ in the manner of Landau and Pollak~\cite{landau} is also based on a similar condition and assumes a similar form as is explained later.

To be more explicit, for the signal with bandwidth $W$, we understand the present uncertainty relation (7) as showing that the {\em shortest possible time interval} $[T]$, which can completely confine the signal, should satisfy $|T|\geq 1/W$ when we use $L_{2}$ norm.
Similarly, any
signal which is completely confined within a time interval $[T]$ has a bandwidth $W$
with $W\geq 1/|T|$.

From the point of view of signal recovery in general, the operation of the bandlimiting procedure of the observed signal is important. We have
\begin{eqnarray}
P_{W}\psi(t)&=&\int dt^{\prime} \int_{[W]} dw e^{-2\pi iw(t-t^{\prime})}\psi(t^{\prime})\nonumber\\
&=&\int dt^{\prime} G(t-t^{\prime}; W)\psi(t^{\prime})
\end{eqnarray}
with
\begin{eqnarray}
 G(t-t^{\prime}; W)=\int_{[W]} dw e^{-2\pi iw(t-t^{\prime})}
\end{eqnarray}
which cannot  average out to zero for $|t-t^{\prime}|<1/|W|$. This shows that
the time dependence of the given signal $\psi(t)$ is smeared to the order of
$\Delta t\sim 1/W$, which is another (and more common) implication of the uncertainty principle.
To recover the original bandlimited signal, one needs to perform the bandlimiting
operation on the measured quantity in one way or another, which will close the missing time interval smaller than $T\leq 1/W$.
For example, in the Nyquist~\cite{nyquist} and Shannon~\cite{shannon}  sampling, one generally measures the bandlimited signal of a fixed $[W]$ with $|W|=W$ by dividing the total time interval into sub-intervals $[T^{\prime}]$ which satisfy $|T^{\prime}|\leq 1/W$. See eq.(22) below. Since this condition $|T^{\prime}|\leq 1/W$ is an inequality, there is some freedom in the choice of $|T^{\prime}|$ and, in practice, one may introduce the frequencies larger than the original bandwidth $W$, $W^{\prime}\equiv 1/|T^{\prime}|> W$, in the measured data (i.e., oversampling) depending on the choice of $|T^{\prime}|$ which satisfies  $|T^{\prime}|< 1/W$. This oversampling corresponds to the violation of the uncertainty relation (7).  From  a point of view of frequency representation, we need the frequency band not smaller than $W$ to describe the information contained in the frequency bandwidth $W$.  After an elimination of frequency components outside $[W]$ (by a lowpass filter), the original bandlimited signal is reproduced in Shannon-Nyquist sampling.

\section{Recovering missing information}
We next recapitulate the basic mechanism to recover the missing information by following the presentation of Donoho and Stark in~\cite{donoho}:
 A signal $s(t)=\langle t|s\rangle \in L_{2}$
defined in a sufficiently large interval $[T^{0}]$ is transmitted to a receiver who knows that $s(t)$ is bandlimited, meaning that $s(t)$ was synthesized using only frequencies in an interval $[W]$. Equivalently,
\begin{eqnarray}
s_{W}(t)\equiv P_{W}s(t)=s(t),
\end{eqnarray}
where $P_{W}$ is the bandlimiting operator defined by the above projection operator. Now suppose the receiver is unable to observe all of $s_{W}(t)$; a certain sub-interval $[T]$ of $t$-values is unobserved. Moreover, the
observed signal is contaminated by observational noise $n(t)=\langle t|n\rangle \in L_{2}$. Thus the received signal $r(t)=\langle t|r\rangle$ satisfies
\begin{equation}
  r(t) =
  \begin{cases}
    s_{W}(t)+ n(t) &  t\in [T^{C}] \\ 0 & t\in [T],
  \end{cases}
\end{equation}
where $[T^{C}]=[T^{0}]-[T]$ is the complement of the interval $[T]$, and we have assumed (without loss of generality) that $n(t)=0$ on $[T]$. Equivalently,
\begin{eqnarray}
r(t)=(1-P_{T})s_{W}(t)+n(t)
\end{eqnarray}
where $1$ is the identity operator $(1f)(t)=f(t)$ which is given by $P_{T^{0}}$ in our setting of the problem.
The receiver's aim is to reconstruct the transmitted signal $s_{W}(t)$ from the noisy received signal $r(t)$. Although it may seem that information about $s_{W}(t)$ for $t\in [T]$ is completely unavailable, the uncertainty principle says recovery is possible provided $|T||W| < 1$~\cite{donoho}.

The basic idea is to re-write the equation (13) using $P_{W}s_{W}(t)=s_{W}(t)$ as
\begin{eqnarray}
r(t)=(1-P_{T}P_{W})s_{W}(t)+n(t)
\end{eqnarray}
and note that the solution of this equation is unique in the absence of the noise or for a  given fixed noise. Assume that two solutions $s_{1}(t)$ and $s_{2}(t)$ satisfy the above equation, then we have
\begin{eqnarray}
(1-P_{T}P_{W})(s_{1}(t)-s_{2}(t))=0
\end{eqnarray}
which implies $||s_{1}(t)-s_{2}(t)||=||P_{T}P_{W}(s_{1}(t)-s_{2}(t))||\leq
||P_{T}P_{W}||||(s_{1}(t)-s_{2}(t))||<||(s_{1}(t)-s_{2}(t))||$, but this is a contradiction if $||s_{1}(t)-s_{2}(t)||\neq 0$. Here we used the relation (5),
$||P_{T}P_{W}||\leq \sqrt{TW}<1$. By noting the fact that the operator
$1/(1-P_{T}P_{W})$ is well defined for $||P_{T}P_{W}||<1$, we have the unique
solution
\begin{eqnarray}\label{DonohoFormula}
s_{W}(t)&=&\frac{1}{1-P_{T}P_{W}}r(t)\nonumber\\
&=&r(t) +\sum_{k=1}^{\infty}(P_{T}P_{W})^{k}r(t)
\end{eqnarray}
in the absence of the noise $n(t)=0$. The noise is important when one analyzes the stability of the solution~\cite{donoho}.
This (16) is a remarkable formula to recover the original signal precisely from the observed signal $r(t)$ in the interval $[T^{C}]=[T^{0}]-[T]$ and the remaining signal in the interval $[T]$ provided by the second term. This is the basic mechanism of Donoho and Stark~\cite{donoho}.

\section{Measurement and disturbance}
The bandlimited property of the right-hand side  is not manifest  in the formula (16). One can explicitly show that sharp $|T|$ modifies the bandlimit of $r(t)$. By noting $r(t)=(1-P_{T})s_{W}(t)$ in (13) in the absence of the noise $n(t)=0$, one can establish
\begin{eqnarray}
\langle r|P_{W^{c}}|r\rangle=\langle s_{W}|P_{T}P_{W^{c}}P_{T}|s_{W}\rangle\neq 0,
\end{eqnarray}
for $WT<1$ with $P_{W^{c}}=1-P_{W}$ using $P_{W^{c}}|s_{W}\rangle=0$, since
\begin{eqnarray}
\langle s_{W}|P_{T}(1-P_{W})P_{T}|s_{W}\rangle/\langle s_{W}|P_{T}|s_{W}\rangle\geq 1-WT>0
\end{eqnarray}
using the relation (4) with $P_{T}$ and $P_{W}$ interchanged. It is remarkable that the bandlimit of the {\em observed signal} $r(t)$ is modified by the presence of the unobserved short interval, but this is an inevitable consequence of the specification of the missing time interval with $WT<1$. For  $WT\geq 1$ we cannot make a definite statement on the bandlimit of $r(t)$.

The bandlimited property of the signal $s_{W}(t)$ is intrinsic and the interval $[T]$ is an external accidental parameter, and in fact $s_{W}(t)$ is independent of $[T]$ if the signal recovery (16) is perfect. One may thus apply the bandlimiting operator to both sides of (16) to obtain by noting $P_{W}s_{W}(t)=s_{W}(t)$,
\begin{eqnarray}\label{NewFormula}
s_{W}(t)&=&P_{W}r(t) +\sum_{k=1}^{\infty}(P_{W}P_{T})^{k}P_{W}r(t)\nonumber\\
&=&\frac{1}{1-P_{W}P_{T}}P_{W}r(t)
\end{eqnarray}
which shows that the bandlimited signal $s_{W}(t)$ is recovered from the  quantity $P_{W}r(t)$, which is constructed  by bandlimiting the measured $r(t)$ that is originally defined in the domain $[T^{C}]=[T^{0}]-[T]$.
This formula (19) incorporates both of the post-measurement smearing of the missing time interval by bandlimiting in addition to the deterministic recovery by
an inversion of the well-defined operator, and it is equally valid as (16).
 Note that
\begin{eqnarray}
P_{W}r(t)=\int_{[T_{0}]}dt^{\prime}\int_{[W]}dwe^{-2\pi iw(t-t^{\prime})}r(t^{\prime})
\end{eqnarray}
spreads over the entire domain of time $t$ without the missing interval $[T]$ if $WT<1$ since $\int_{[W]}dwe^{-2\pi iw(t-t^{\prime})}$ cannot average out to zero for $|t-t^{\prime}|< 1/W$, as was explained in (10). This shows that the missing time interval $[T]$ is closed if the bandlimit is imposed on the observed data, and thus even the first term in (19) has no missing time interval although it does not completely recover the original signal by itself.

We suggest the formula (19) as an alternative to the original Donoho-Stark formula (16). The relations (18) and (19) show that if the unobserved time interval is very small, $WT \ll 1$, the band limit of $r(t)$ is significantly modified and goes far beyond the original $[W]$; in such a case, by recalling the relation $||P_{T}P_{W}||\leq \sqrt{TW}\ll 1$, the first term of (19), namely, a simple bandlimiting of the observed signal will provide a good approximation to the original signal. In practical applications, a detector with the time resolution of $10^{-8}$ sec, for example, cannot exclude  unobserved short time intervals such as $10^{-12}$ sec and one cannot recover all of those (infinitely many) short unobserved intervals by (16). The first term in the modified formula (19) automatically takes care of such short unobserved intervals by  smearing the signals as in (20) for $WT \ll 1$. This will be numerically illustrated later.

It is generally assumed in {\em classical physics} that observation does not modify (or destroy) the signal. Consequently, it is assumed that, in principle, no limit to the accuracy in the time or frequency resolution of the detector.
However, our analysis of (18) shows that the obtained signal is significantly modified by precise measurements or by the identification of a short unobserved time interval. We discuss this issue in the following.

To detect the missing
short time interval $[T]$ in $r(t)$, one needs to measure the time dependence of the observed
$r(t)$ with corresponding accuracy. Moreover, one needs to ensure that the observed data satisfies $r(t)=s_{W}(t)$, namely, bandlimited  for all the time $t$ except for the interval $[T]$. For simplicity, we assume the vanishing noise $n(t)=0$. Our basic assumption is to describe the given signal by the projective measurements $\langle s_{W}|P_{T^{\prime}}|s_{W}\rangle=\int_{[T^{\prime}]}dt|s_{W}(t)|^{2}$ and
$\langle s_{W}|P_{W^{\prime}}|s_{W}\rangle=\int_{[W^{\prime}]}dw|\hat{s}_{W}(w)|^{2}$ which are consistent with our use of $L_{2}$ norm.  Starting with the {\em observed} signal $r(t)$, one may divide the total time $[T^{0}]-[T]$ into small sub-intervals specified by critical $[T_{c}]$ and examine the assumed relation $r(t)=s_{W}(t)$ projectively in each interval $[T_{c}]$, $\langle r|P_{T_{c}}|r\rangle=\int_{[T_{c}]}dt |r(t)|^{2}$, where the different center of each interval is implicit; we examine each segment of the obtained signal $P_{T_{c}}|r\rangle/||P_{T_{c}}|r\rangle||$
and check if this segment is bandlimited within $W$,
\begin{eqnarray}
\langle r|P_{T_{c}}P_{W}P_{T_{c}}|r\rangle/\langle r|P_{T_{c}}|r\rangle=1.
\end{eqnarray}
Using the bound to the conditional measurement in (4) with $P_{T}$ and $P_{W}$ interchanged, one then obtains a necessary condition
$|T_{c}|W\geq 1$.
This relation, in particular $|T_{c}|W = 1$, gives a condition to obtain the {\em reliable and sufficient information} to describe the classical signal by the projective analyses $P_{W}$ and $P_{T_{c}}$. This is precisely what the uncertainty relation (7) tells. In the present formulation, classical and quantum measurements become rather similar.

The identification of the unobserved short interval $[T]$ with $TW<1$ thus inevitably disturbs
the bandlimited property  of the {\em obtained data}, as (18) indicates. Donoho-Stark mechanism and also Shannon-Nyquist sampling, which is explained in further detail later, allow the significant modification of the obtained  signal by the identification of the unobserved short interval or by the active measurements of short intervals, but it is {\em assumed} that one can later recover the original signal from the observed data.
A crucial difference from quantum mechanics is that we do not have the notion of reduction in classical physics; the conditional measurement in (4) uses the modified state for the second measurement but the original state is implicitly assumed to be still there even after the first measurement. The {\em obtained data} are modified, namely, the observed shape of the signal is different from the original one but the original signal is still there as is seen in $|r(t)\rangle=(1-P_{T})|s(t)\rangle$ and  $s(t)$ is recovered by inversion; the uniqueness proof in (15) depends on the same bandlimited property of the original signal  $s(t)$ even after the measurement. In contrast, reduction implies that the initial state disappears after the  measurement in quantum mechanics.

As for the general  recovery of missing time intervals, a missing interval $[T]$ with $TW > 1$ is fatal to Donoho-Stark mechanism (and also to Shannon-Nyquist sampling) since the bulk of the information may be lost or mathematically $1-P_{T}$ is not inverted. The recovery of such missing information is "ill-posed" in the conventional sense, and this is precisely where the  compressed sensing scheme works with the aid of a priori information such as sparsity and low-rank
assumptions~\cite{candes1, donoho2, candes2}.

\section{Shannon-Nyquist sampling and Donoho-Stark mechanism}

We have discussed  a modified  Donoho-Stark formula (19),
in place of the original one (16), which incorporates both of the post-measurement smearing of the unobserved interval $[T]$ in $r(t)$ by bandlimiting operation and the  deterministic recovery.  
This property suggests the common basis of  Shannon-Nyquist sampling, which restores the original signal by combining a sampling of  short time intervals $T\leq 1/W$ with suitable later
bandlimiting operation using Fourier analysis, and  Donoho-Stark mechanism, which
restores the signal in a specific short interval $T<1/W$ with the help of the uncertainty principle; Fourier analysis and the uncertainty principle are closely related in classical physics.

To be more specific, it is natural to assume that one knows the values of the signal at both ends of the unobserved time interval, $T_{k}$ and $T_{k+1}$
with $T_{k+1}-T_{k}=T$, in the analysis of Donoho and Stark since they assume that the entire signal was precisely measured outside the specific interval $[T]$. See (12) with a vanishing noise $n(t)=0$. See also Fig.1. To be exact, one needs to define the unobserved interval by  $[T]=\{t: T_{k}+\epsilon/2 \leq t\leq T_{k+1}-\epsilon/2\}$ with an infinitesimal positive $\epsilon$ but we forgo the technical details. One may then divide the observed signal into equal sub-intervals with a size $T$ and denote the end points of those intervals by $\{T_{0}, ..., T_{k-1}\}$ and $\{T_{k+2}, ...,T_{N}\}$ with $N$ a minimum integer which satisfies $T_{N}- T_{0}\geq |T^{0}|$, where $[T^{0}]$ stands for the total time interval in which the entire signal is contained. Since one knows $s(kT)$ for $k=0, ...,N$ and $s(kT)=0$ for all other $k$, one can apply
the Shannon-Nyquist reconstruction formula
\begin{eqnarray}
s_{W^{\prime}}(t)=\sum_{k\in Z}s(kT)h_{T}(t-kT)
\end{eqnarray}
where $W^{\prime}\equiv 1/T$ and the sinc-function $h_{T}(t)=\sin(\pi t/T)/(\pi t/T)$ which forms a complete orthonormal set $\int_{-\infty}^{\infty}h_{T}(t-kT)h_{T}(t-k^{\prime}T)dt/T=\delta_{k,k^{\prime}}$; this formula describes $s_{W^{\prime}}(t)$ which contains the frequency $-\frac{1}{2T}\leq w \leq \frac{1}{2T}$ in the Fourier representation and converges in the sense
of $L_{2}$ norm~\cite{text}.  

Since $1/T = W^{\prime}>W$,
this signal corresponds to oversampling (and this signal corresponds to the detection of the domain below the uncertainty limit from a point of view of the uncertainty principle), and thus one may apply bandlimiting operation down to $W$ (or a lowpass filter) to recover the original signal $s_{W}(t)$. To show this, we define a discretized {\em measured} signal $|r\rangle_{SN}$ in Shannon-Nyquist sampling by
\begin{eqnarray}
|s\rangle=\int dt|t\rangle\langle t|s\rangle \Rightarrow |r\rangle_{SN}\equiv\sum_{k\in Z}T|kT\rangle\langle kT|s\rangle,
\end{eqnarray}
which corresponds to a sampled signal $r_{SN}(t)=\langle t|r\rangle_{SN}=\sum_{k\in Z}T\delta(t-kT)\langle kT|s\rangle$,
and one can confirm that $s_{W^{\prime}}(t)$ in (22) is given by $s_{W^{\prime}}(t)=\langle t|P_{W^{\prime}}|r\rangle_{SN}$. More generally,
\begin{eqnarray}
s_{W}(t)&\equiv&\langle t|P_{W}|r\rangle_{SN}\nonumber\\
&=&\sum_{k\in Z}T\langle t|P_{W}|kT\rangle\langle kT|s\rangle\nonumber\\
&=&TW\sum_{k\in Z}\frac{\sin \pi W(t-kT)}{\pi W(t-kT)}s(kT)
\end{eqnarray}
where we defined $s(kT)=\langle kT|s\rangle$. This $s_{W}(t)$ is well-defined for
$W\leq W^{\prime}$ since $P_{W}s_{W^{\prime}}(t)=\langle t|P_{W}P_{W^{\prime}}|r\rangle_{SN}=\langle t|P_{W}|r\rangle_{SN}=s_{W}(t)$ by noting $P_{W}P_{W^{\prime}}=P_{W}$ if one chooses $[W]\subseteq [W^{\prime}]$. Namely, $s_{W}(t)$ is a bandlimited version of $s_{W^{\prime}}(t)$ defined by the Shannon-Nyquist reconstruction formula (22) and agrees with the original bandlimited signal.

We have $s_{W^{\prime}}(kT)=s(kT)$ in (22), but this property is not explicit for $s_{W}(t)$ in (24). To clarify this issue, we use the relation,
\begin{eqnarray}
\langle w|r\rangle_{SN}&=&\sum_{k\in Z}T\langle w|kT\rangle\langle kT|s\rangle\nonumber\\
&=&\sum_{k\in Z}T e^{i 2 \pi w kT}s(kT)=\sum_{k\in Z}\hat{s}(w-\frac{k}{T})
\end{eqnarray}
where we used Poisson summation formula at the last step with $\hat{s}(w) = \int_{-\infty}^{+\infty}dt s(t)e^{i 2 \pi w t}$. Thus, the Shannon-Nyquist sampling process leads to a {\em periodization} of the Fourier transform of $s(t)$. We are considering a bandlimited $s(t)$, namely, $\hat{s}(w)$ has support in $[W]$. We can then avoid aliasing (i.e., the overlap of adjacent terms in the last expression in (25)) if we choose  $T \leq 1/W$. The set of values $\{s(kT)\}$ cannot be arbitrary and they are constrained by (25). For such $\{s(kT)\}$, we have $\hat{s}(w)=\langle w|r\rangle_{SN}$ for $w\in [W]$ since only the term with $k=0$ survives for $w\in [W]$ in the last expression in (25). We thus have $s(t)=s_{W}(t)$ for $s_{W}(t)$ defined in (24).

We have recapitulated the basic procedure of Shannon-Nyquist sampling by incorporating the idea of the uncertainty principle emphasized by Donoho and Stark. Since $\{r(kT)\}=\{s(kT)\}$ by assumption, the knowledge of the observed $r(t)$ is sufficient to reproduce $s_{W}(t)$, and thus the signal recovery of Donoho and Stark is justified by Shannon-Nyquist sampling theorem;
the difference is that Donoho-Stark gives the precise result while Shannon-Nyquist needs a lowpass filter to recover the original signal. The connection between Shannon-Nyquist sampling and Donoh-Stark mechanism is schematically shown in Fig.1.
\\

\begin{figure}
\begin{center}
\begin{tabular}{c}
\includegraphics[height=4cm] {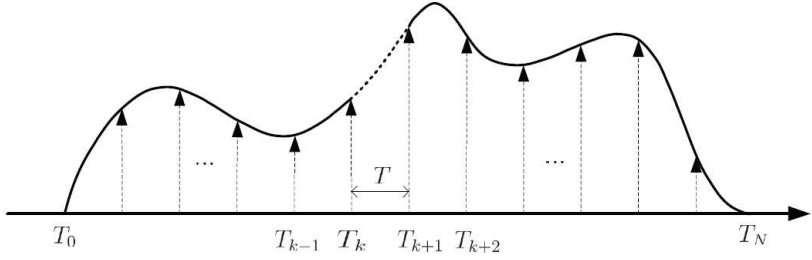}
\end{tabular}
\end{center}
\caption{Schematic figure for the connection between Shannon-Nyquist sampling and Donoho-Stark mechanism.}
\label{Fig1}
\end{figure}

\noindent {\bf A new signal recovery formula}\\
\\  
We now illustrate a new scheme of the information recovery from the domain below the uncertainty limit using the idea of Shannon-Nyquist sampling by taking the shape of $r(t)$ into account. We thus define
a Shannon-Nyquist sampled signal corresponding to the measured value $r(t)$ in Donoho-Stark mechanism,
$r_{DS}(t)\equiv \sum_{k\in Z}T_{SN}\delta(t-kT_{SN}) r(kT_{SN})$,
where $|r\rangle=(1-P_{T_{DS}})|s\rangle$. To treat a general case, we introduced
two time intervals, Shannon-Nyquist $T_{SN}$ and Donoho-Stark $T_{DS}$, which satisfy $T_{DS}< T_{SN}\leq 1/W$.  We then have the Poisson summation formula
\begin{eqnarray}
\sum_{k\in Z}T_{SN} e^{i 2 \pi w kT_{SN}}r(kT_{SN})=\sum_{k\in Z}\hat{r}(w-\frac{k}{T_{SN}}).
\end{eqnarray}
From (25) and (26), we obtain the relation
\begin{eqnarray}
\sum_{k\in Z}\hat{s}(w-\frac{k}{T_{SN}})=\sum_{k\in Z}\hat{r}(w-\frac{k}{T_{SN}})
\end{eqnarray}
since $s(kT_{SN})=r(kT_{SN})$.

By restricting $w\in [W]$ in (27),  we have
\begin{eqnarray}
\hat{s}(w)&=&P_{W}\hat{r}(w)+
\sum_{k=1}^{\infty}P_{W}[\hat{r}(w-\frac{k}{T_{SN}})+\hat{r}(w+\frac{k}{T_{SN}})]
\end{eqnarray}
for $W\leq 1/T_{SN}$ since only the term with $k=0$ on the left-hand side of (27) is non-vanishing for $w\in [W]$; note that $\hat{s}(w)$ is non-vanishing only for $w\in [W]$. The knowledge of $\hat{r}(w)=\langle w|r\rangle=\langle w|(1-P_{T_{DS}})|s\rangle$ for
$w\in (-\infty, \infty)$ is thus sufficient to recover $\hat{s}(w)$ bandlimited in $w\in [W]$ and thus original $s(t)$. Note that $T_{DS}< T_{SN}\leq 1/W$.
 The basis for the above relation (28) is that $s(kT_{SN})=r(kT_{SN})$ for all $k$ but the bandlimit is different for $\hat{s}(w)$ and $\hat{r}(w)$ due to the operation $P_{T_{DS}}$. See (18).

The relation (28) is analogous to (19), but there exists a  difference.
All the terms on the right-hand side are expressed by $P_{W}\hat{r}(w)$ in Donoho-Stark formula (19), while the extended bandwidth of $\hat{r}(w)$ beyond $1/T_{SN}$ is crucial in (28), which is analogous to the original Donoho-Stark formula (16) with the extended bandwidth of $r(t)$ as in (17). In retrospect, (28) is valid for $r(t)$ of  {\em any shape} with $r(kT_{SN})=s(kT_{SN})$ and extended bandwith beyond $1/T_{SN}$, not necessarily the form in Donoho-Stark mechanism. In all those cases the observed $r(t)$ reproduces the bandlimited $s(t)$, but at the same time the refined (below the uncertainty limit) time variation of $r(t)$  is generally lost in this procedure. This aspect of smearing the information below the uncertainty limit
is another important aspect of our information recovery formula (28), and it is close to our common understanding of the quantum mechanical uncertainty relation.
\\

\begin{figure}
\begin{center}
\begin{tabular}{c}
\includegraphics[height=9cm] {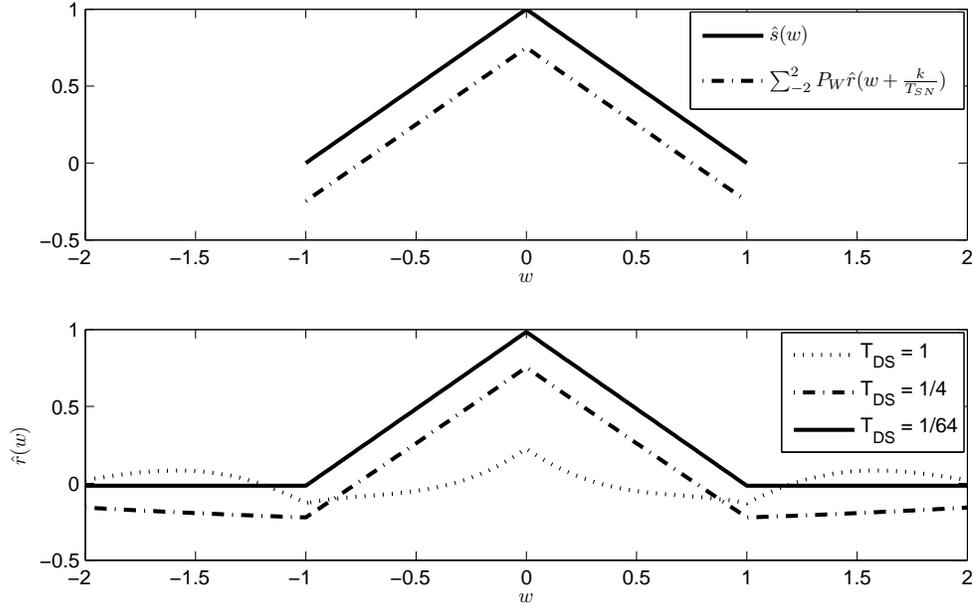}
\end{tabular}
\end{center}
\caption{In the upper graph, the original signal $\hat{s}(w)$ and the prediction of (28) with the first 5 terms for $T_{DS}=T_{SN}=1/4s$ are shown. In the lower graph, the Fourier transform of the observed signal $\hat{r}(w)$ in Donoho-Stark mechanism (with $W=2Hz$ and $T_{DS}=1s, 1/4s, 1/64s$) is shown. As $T_{DS}W$ decreases, $P_W \hat{r}(w)$ approaches $\hat{s}(w)$ given in the upper graph. When $T_{DS}W > 1$ (with $T_{DS}=1s$), $P_W \hat{r}(w)$ is seriously distorted and can not be used to recover the original signal via \eqref{NewFormula} (since ($1-P_WP_T$) is non-invertible in this case).}
\label{Fig2}
\end{figure}

\noindent {\bf Numerical illustration}\\
\\
Finally, we show that the bandlimited first term in Donoho-Stark mechanism (19), where $|r\rangle=(1-P_{T_{DS}})|s\rangle$, gives a useful approximation to $s(t)$.
This definition of $|r\rangle$ is written as
\begin{eqnarray}
\langle w|r\rangle
&=&\langle w|s\rangle-T_{DS}\int_{[W]}dw^{\prime}\frac{\sin \pi T_{DS}(w-w^{\prime})}{ \pi T_{DS}(w-w^{\prime})}\langle w^{\prime}|s\rangle
\end{eqnarray}
where we set $t_{0}=0$ in the definition of $P_{T_{DS}}$ in (1).
The formula (19) corresponds to an iterative solution to the "integral equation" (29) defined for $w\in [W]$ by treating $T_{DS}$ as a small
coupling constant.  For small $T_{DS}W \ll 1$, one can show that the bandlimited $P_W \hat{r}(w)=P_W\langle w|r\rangle$ provides a very good approximation to the original  $\hat{s}(w)$, namely, (29) gives
\begin{eqnarray}
P_W \hat{r}(w)\simeq \hat{s}(w)- WT_{DS}(\int_{[W]}dw^{\prime}\hat{s}(w^{\prime})/W),
\end{eqnarray}
which is illustrated  in Fig. \ref{Fig2} by assuming a specific example of $\hat{s}(w)$ with $1/W=1/2$; in $t$-representation, this $\hat{s}(w)$ corresponds to $s(t)=2(1-\cos2\pi t)/(2\pi t)^{2}$ which is non-negative and assumes
$s(0)=1$, $s(\pm 1/2)=4/\pi^{2}\sim 1/2$ and $s(\pm 1)=0$.

We also illustrate our proposed signal recovery formula (28) for the case $T_{DS}=T_{SN}=1/4<1/W=1/2$, namely, the recovery of the signal, for which the peak values of the signal $s(t)$ between $s(1/8)$ and $s(-1/8)$ is missing, is shown by a dash-dot figure in the upper graph in Fig.2 ($T_{DS}=T_{SN}$ is allowed since both are smaller than $1/W$). The first term $P_W \hat{r}(w)$ alone almost recovers the signal, but the convergence of the rest of terms is rather slow, which we have confirmed by summing the series up to first  5 terms with $k=0, \pm1,\pm2$ in (28).  This slow convergence is partly related to the sharp cut-off induced by $P_{T_{DS}}$. This use of the  first term $P_W \hat{r}(w)$ alone is close to the customary procedure to deal with the missing time domain of sub-uncertainty limit in quantum mechanics using only the information allowed by the detector capacity.

The analysis in this section shows  that the original band-limited signal is, in principle, recovered without knowing $s(t)$ within
an interval $[T]$ below the uncertainty limit $T<1/W$ by a variety of ways. Donoho and Stark have given an attractive  physical picture, namely, the uncertainty principle, for Shannon-Nyquist sampling and related formulas which are often discussed using the Poisson summation formula~\cite{note}.

\section{Recovery of missing states in quantum mechanics} 

It is interesting to examine the possible information recovery below the uncertainty limit in quantum mechanics. 
From a point of view of the uncertainty relation, the momentum-coordinate uncertainty relation in the manner of Landau-Pollak~\cite{landau} is close to the uncertainty relation used by Donoho and Stark in information theory, although the notion of reduction is crucial in Landau-Pollak-type uncertainty relation. We thus consider the probability amplitude $\psi(t,x)=\exp[-2\pi iEt]\psi(x)$ in this paper and discuss a possible recovery of an unobserved small interval $[X]$ in $\psi(x)$ for which allowed momentum is limited.

We first recall the definition of Landau-Pollak-type uncertainty relation. The projection operators are defined by
\begin{eqnarray}
P_{P}=\int_{p_{0}-\frac{1}{2}P}^{p_{0}+\frac{1}{2}P}dp|p\rangle\langle p|,\ \ \
P_{X}=\int_{x_{0}-\frac{1}{2}X}^{x_{0}+\frac{1}{2}X}dx |x\rangle\langle x|,
\end{eqnarray}
using the relations $\langle x|x^{\prime}\rangle=\delta(x-x^{\prime}),\
\langle p|p^{\prime}\rangle=\delta(p-p^{\prime}), {\rm and}\
\langle x|p\rangle=\exp[2\pi ipx]$
where we adopt the convention $2\pi\hbar=1$. The choice $p_{0}=0$ is convenient in our application.
As for the uncertainty principle, the upper bound to the probability of conditional measurement (4) is now replaced by
\begin{eqnarray}
\frac{\langle\psi|P_{P}P_{X}P_{P}|\psi\rangle}{\langle\psi|P_{P}|\psi\rangle}
&=&\frac{\langle\psi|P_{P}(P_{P}P_{X}P_{P})P_{P}|\psi\rangle}{\langle\psi|P_{P}P_{P}|\psi\rangle}
\nonumber\\
&\leq& ||P_{P}P_{X}P_{P}||=||P_{X}P_{P}||^{2}
\leq PX,
\end{eqnarray}
and the condition for the consistent description of a state in terms of $P_{X}$ and $P_{P}$ (or compatibility of $P_{X}$ and $P_{P}$), $\langle\psi|P_{P}P_{X}P_{P}|\psi\rangle/\langle\psi|P_{P}|\psi\rangle=1$, implies
\begin{eqnarray}
XP\geq 1,
\end{eqnarray}
which is the  Landau-Pollak-type uncertainty relation~\cite{landau}.
We here comment on a specific aspect of the  quantum mechanical state recovery related to reduction.
The notion of reduction implies that we have 
\begin{eqnarray}
|\psi_{P}\rangle \rightarrow |\psi_{M}\rangle=(1-P_{X})|\psi_{P}\rangle/||(1-P_{X})|\psi_{P}\rangle||
\end{eqnarray}
after the measurement of $(1-P_{X})$ in quantum mechanics, while we have 
\begin{eqnarray}
|\psi_{P}\rangle \rightarrow |\psi_{P}\rangle=(1-P_{X})|\psi_{P}\rangle+P_{X}|\psi_{P}\rangle
\end{eqnarray}
after the measurement of $(1-P_{X})$ in classical theory, namely, the state itself is not modified by measurement. Thus the recovery of
the original state $|\psi_{P}\rangle$ is natural in classical theory but the quantum case is conceptually more involved.

The {\em deterministic} state recovery from the measured data of coordinate in one-dimensional quantum mechanical problem is 
known~\cite{bertrand,raymer,janicke,leonhardt1,leonhardt2}, and  it is used to analyze the possible recovery of a small missing interval $[X]$  in the coordinate with $XP<1$ by analyzing a  prepared state 
\begin{eqnarray}
|\psi_{M}\rangle=(1-P_{X})|\psi_{P}\rangle/||(1-P_{X})|\psi_{P}\rangle||
\end{eqnarray}
where the momentum is initially limited within $[P]$ for the state $|\psi_{P}\rangle$, namely, $P_{P}|\psi_{P}\rangle=|\psi_{P}\rangle$. Our purpose is to
recover the state $|\psi_{P}\rangle$ from the given $|\psi_{M}\rangle$ for $XP<1$. 

The analysis of probabilities shows
\begin{eqnarray}
\frac{\langle\psi_{P}|P_{X}P_{P}P_{X}|\psi_{P}\rangle}{\langle\psi_{P}|P_{X}|\psi_{P}\rangle}&<&XP < 1,\nonumber\\
\frac{\langle\psi_{P}|(1-P_{X})P_{P^{c}}(1-P_{X})|\psi_{P}\rangle}{\langle\psi_{P}|P_{X}|\psi_{P}\rangle}&=&1-\frac{\langle\psi_{P}|P_{X}P_{P}P_{X}|\psi_{P}\rangle}{\langle\psi_{P}|P_{X}|\psi_{P}\rangle}\nonumber\\ &>& 1-XP>0,
\end{eqnarray}
and thus both $P_{X}|\psi_{P}\rangle$ and $(1-P_{X})|\psi_{P}\rangle$ contain momenta outside the momentum-limit $|P|$.
Here we defined $P_{P^{c}}\equiv 1-P_{P}$ which satisfies $P_{P^{c}}P_{P}=0$ and $P_{P^{c}}|\psi_{P}\rangle=0$. 
The prepared state $|\psi_{M}\rangle$, which has a gap in the coordinate dependence, thus spoils the momentum-limited property.  
It is conceptually simpler to discuss the recovery of $|\psi_{P}\rangle$ from 
\begin{eqnarray}
P_{P}|\psi_{M}\rangle=(1-P_{P}P_{X}P_{P})|\psi_{P}\rangle/||(1-P_{P}P_{X}P_{P})|\psi_{P}\rangle||
\end{eqnarray}
which has no gap in coordinate space
\begin{eqnarray}
&&\langle x|P_{P}|\psi_{M}\rangle\times||(1-P_{P}P_{X}P_{P})|\psi_{P}\rangle||\nonumber\\
&&=\langle x|\psi_{P}\rangle
-\int_{P}dp \int_{X} dy e^{2\pi ip(x-y)}\langle y|\psi_{P}\rangle\nonumber\\
&&=\langle x|\psi_{P}\rangle - XP
\int_{X} \frac{dy}{X}\frac{\sin \pi P(x-y)}{\pi P(x-y)}\langle y|\psi_{P}\rangle
\end{eqnarray}
where $0<XP\frac{\sin \pi P(x-y)}{\pi P(x-y)}<1$ for $x, y\in X$ with $XP<1$. Thus the gap in $\langle x|\psi_{M}\rangle$ for $x \in X$ is smoothed and disappears in $\langle x|P_{P}|\psi_{M}\rangle$  for $x \in X$ with $XP<1$, which is regarded as a consequence of the ordinary uncertainty principle in quantum mechanics. The use of $P_{P}|\psi_{M}\rangle$ corresponds to the adoption of the  modified version of Donoho-Stark scheme discussed in (19) and (20).

We now sketch the basic idea and procedure of the deterministic reconstruction of the quantum mechanical state $P_{P}|\psi_{M}\rangle$, which is momentum-limited and has no gap in coordinate, from the measured data following the formulation of Leonhardt and Schneider~\cite{leonhardt2}, which is based on the Hamiltonian $\hat{H}=\frac{\hat{p}^{2}}{2m}+U(x)$ with an arbitrary stationary potential. Only the case of a free Hamiltonian with $U(x)=0$ is discussed in the present paper, for simplicity. They start
with the time dependent density matrix
\begin{eqnarray}
\rho_{M}(t)\equiv e^{-i\hat{H}t}P_{P}|\psi_{M}\rangle\langle \psi_{M}|P_{P}e^{i\hat{H}t}.
\end{eqnarray}
After the {\em assumed ideal measurements} of $x$-dependence by the projection operator $P_{x}=|x\rangle\langle x|$, one obtains the diagonal elements of the density matrix 
\begin{eqnarray}
\rho_{f}(t)&=&\sum_{x}|x\rangle\langle x|e^{-i\hat{H}t}\rho_{M}(0)e^{i\hat{H}t}|x\rangle\langle x|\nonumber\\
&=&\sum_{x}\sum_{p, p^{\prime}\in P}|x\rangle\langle x|p\rangle e^{-i(\omega(p)-\omega(p^{\prime}))t}\langle p|\rho_{M}(0)|p^{\prime}\rangle\langle p^{\prime}|x\rangle\langle x|
\end{eqnarray}
where $\langle p|e^{-i\hat{H}t}=\langle p|e^{-i\omega(p)t}$ with $\omega(p)$ standing for the kinetic energy of the particle.
The appearance of only the diagonal elements is a result of quantum mechanical reduction, and the important idea in their analysis~\cite{leonhardt1,leonhardt2} is the examination of time dependence in (41) which supplies extra information not available by the measurements of spatial dependence; one can thus determine those {\em off-diagonal elements} specified by $\omega(p)-\omega(p^{\prime})$ from the diagonal elements of the density matrix in the coordinate representation, and one eventually recovers the state in (38), namely, $\langle x|P_{P}|\psi_{M}\rangle=\langle x|(1-P_{P}P_{X}P_{P})|\psi_{P}\rangle/||(1-P_{P}P_{X}P_{P})|\psi_{P}\rangle||$ from the measured data. Note that we need the full density matrix including off-diagonal elements to determine each state contained in the density matrix. Further details are found in~\cite{leonhardt2}.

One may thus recover the original state $|\psi_{P}\rangle$ by an inversion operation applied to $P_{P}|\psi_{M}\rangle$ by noting that the non-negative hermitian operator $P_{P}P_{X}P_{P}$ satisfies $||P_{P}P_{X}P_{P}||\leq XP<1$ in (32), and 
\begin{eqnarray}
|\psi_{P}\rangle&=&\frac{1}{1-P_{P}P_{X}P_{P}}P_{P}|\psi_{M}\rangle||(1-P_{P}P_{X}P_{P})|\psi_{P}\rangle||\nonumber\\
&=&[1+P_{P}P_{X}P_{P}+(P_{P}P_{X}P_{P})^{2}+ ...]P_{P}|\psi_{M}\rangle||(1-P_{P}P_{X}P_{P})|\psi_{P}\rangle||,
\end{eqnarray}
which is analogous to the procedure of Donoho and Stark. The factor $||(1-P_{P}P_{X}P_{P})|\psi_{P}\rangle||$ may be treated as a normalization constant to be fixed after solving for $|\psi_{P}\rangle$. Alternatively, one may solve the  equation (39) in $x\in X$, which is regarded as a well-defined integral equation, for a small coupling $XP$ iteratively for a {\em given} $\langle x|P_{P}|\psi_{M}\rangle$  and determine $\langle x|\psi_{P}\rangle$ for $x\in X$, which is the part of the state to be recovered. Originally, eq.(39) was defined to study the coordinate dependence of $\langle x|P_{P}|\psi_{M}\rangle$, but now regarded as a functional relation between $\langle x|\psi_{P}\rangle$ and $\langle x|P_{P}|\psi_{M}\rangle$. The analysis we performed so far is, given the state $\langle x|P_{P}|\psi_{M}\rangle$, how to measure it and how to infer the state  $\langle x|\psi_{P}\rangle$.

An alternative formulation may be to measure a given momentum-limited state 
$|\psi_{P}\rangle$ but missed the measurement of its tiny part $P_{X}|\psi_{P}\rangle$ and thus the recovered state in the analysis of Ref.~\cite{leonhardt2} corresponds to $(1-P_{X})|\psi_{P}\rangle$. In this case, it is natural to assume that one knows $\langle x|(1-P_{X})|\psi_{P}\rangle$. One may then apply the post measurement momentum-limiting operation to obtain 
$\langle x|P_{P}(1-P_{X})|\psi_{P}\rangle=\langle x|(1-P_{P}P_{X}P_{P})|\psi_{P}\rangle$, which closes the gap $P_{X}|\psi_{P}\rangle$ in the coordinate dependence, and apply the inversion procedure discussed above to recover $\langle x|\psi_{P}\rangle$. 

It remains to be seen if the state recovery procedures we discussed are useful in the practical settings of quantum mechanical problems. The recovery of the lost information in the domain $XP<1$ means that the loss of information in such a small domain is not fatal, which is in accord with our common understanding of the uncertainty principle, although its precise recovery is something we are not used to in quantum mechanics. 
  
As another interesting quantum mechanical problem, one may analyze the time-energy uncertainty relation which is less precisely defined compared to the momentum-coordinate uncertainty relation~\cite{ahanonov-bohm, busch} and not strictly constrained by the notion of reduction ; the reduction of the state by the measurement of energy is well-defined but the reduction of the state due to the measurement of time is not defined in quantum mechanics. It may thus be interesting to examine the possible information recovery from the time interval below the uncertainty limit as in the classical Donoho-Stark mechanism by considering a state such as $\psi(t,x)=\int_{W}dw e^{-iwt}\psi(w,x)$ which is not the eigenstate of energy and thus not stationary. Although the $L_{2}$ norm for the time dependence is not usually adopted in quantum mechanics and thus differs from the case of Donoho-Stark analysis, one may define the relation analogous to $TW\geq 1$ in (7) as a compatibility condition of energy and time measurements in Fourier analysis.

\section{ Discussion and conclusion}
We have analyzed the physical picture behind the recovery of signals from a domain below the uncertainty limit in classical information theory.  We have shown that the Shannon-Nyquist sampling theorem, which is fundamental in signal processing, utilizes essentially the same mechanism as the scheme of Donoho-Stark. The uncertainty principle provides a criterion of Shannon-Nyquist sampling and the specific Donoho-Stark scheme is not regarded as compressed sensing. A new signal recovery formula (28), which is analogous to  Donoho-Stark formula but based on the idea of Shannon-Nyquist sampling, has been given; it illustrates the smearing as well as recovery of information below the uncertainty limit.

We have also discussed the recovery of states from the domain below the uncertainty limit of coordinate and momentum in quantum mechanics and shown that in principle the state recovery, if suitably formulated, works by assuming ideal measurement procedures. Practical aspects of this state recovery remain to be clarified.

One of the important implications of the present analysis is that the uncertainty principle provides a universal sampling criterion covering the classical Shannon-Nyquist sampling theorem and the quantum mechanical measurement, since the general measurement limit in quantum mechanics is set by uncertainty relations.

We have concentrated on the deterministic information recovery in the present paper, but as a related problem which utilizes the compressed sensing, we mention a recent interesting experiment in which weak measurement and compressed sensing were used to measure complementary observables simultaneously. The momentum distribution is directly imaged, while the position distribution is recovered using (classical) compressive sensing in such a manner that the uncertainty principle in quantum mechanics is preserved~\cite{howland}.

As for the conditional measurements in the phase space with $PX<1$, they have also been discussed from a different point of view~\cite{fujikawa} to account for an apparent violation of uncertainty relations in some specific measurement procedures~\cite{ballentine}.
\\

\noindent{\bf Acknowledgments}\\

We thank Hong Lei and Zhifeng Lv for useful comments. One of the authors (K.F.) thanks the hospitality at School of Physics, Beijing Institute of Technology. This work is supported in part by Natural Science Foundation of China (Grant Nos. 11275024 and 61301188), Ministry of Science and Technology, China (Grant No. 2013YQ03059503), and JSPS KAKENHI (Grant No. 25400415).
\\


\begin{thebibliography}{99}
\bibitem{note}
Analogous example in mathematical physics may be the characterization of central charges in Kac-Moody and Virasoro algebras, which are expressed in terms of infinite series, by the notion of quantum anomalies.
\end{thebibliography}
\end{document}